\documentclass{aastex63}

\pdfoutput=1

\newcommand{\Fcm}{F_{10.7}}
\newcommand{\Bfd}{|B|}
\newcommand{\GHz}{\mathrm{GHz}}
\newcommand{\Rt}{R_\theta}
\newcommand{\Req}{R_{212}^{equat.}}
\newcommand{\Rpo}{R_{212}^{polar}}
\newcommand{\Reqq}{R_{405}^{equat.}}
\newcommand{\Rpoo}{R_{405}^{polar}}

\usepackage{multirow}
    \newcommand{\mr}{\multirow}
\newcommand{\dgr}[1]{${#1}^{\circ}$}
\newcommand{\amin}[1]{${#1}^{\prime}$}
\newcommand{\asec}[1]{${#1}^{\prime\prime}$}
\newcommand{\dgrms}[3][2]{\dgr{#1}\amin{#2}\asec{#3}}

\received{}
\revised{}
\accepted{}
\submitjournal{ApJ}

\shorttitle{The subterahertz solar cycle}
\shortauthors{Menezes et al.}

\graphicspath{{./}{figures/}}

\begin{document}

\title{The subterahertz solar cycle: Polar and equatorial radii derived from SST and ALMA}

\correspondingauthor{Fabian Menezes}
\email{menezes.astroph@gmail.com}

\author[0000-0002-4703-4027]{Fabian Menezes}
\affiliation{Centro de R{\'a}dio Astronomia e Astrof{\'i}sica Mackenzie (CRAAM),
             Universidade Presbiteriana Mackenzie,
             S{\~a}o Paulo, Brazil}

\author[0000-0002-5897-5236]{Caius L. Selhorst}
\affiliation{N{\'u}cleo de Astrof{\'\i}sica,
             Universidade Cruzeiro do Sul / Universidade Cidade de S{\~a}o Paulo,
             S{\~a}o Paulo, SP,  Brazil}

\author[0000-0002-8979-3582]{Carlos Guillermo Giménez de Castro}
\affiliation{Centro de R{\'a}dio Astronomia e Astrof{\'i}sica Mackenzie (CRAAM),
             Universidade Presbiteriana Mackenzie,
             S{\~a}o Paulo, Brazil}
\affiliation{Instituto de Astronomía y Física del Espacio,
             UBA/CONICET, Buenos Aires, Argentina.}

\author[0000-0002-1671-8370]{Adriana Valio}
\affiliation{Centro de R{\'a}dio Astronomia e Astrof{\'i}sica Mackenzie (CRAAM),
             Universidade Presbiteriana Mackenzie,
             S{\~a}o Paulo, Brazil}

\begin{abstract}
At subterahertz frequencies -- \textit{i.e.}, millimeter and submillimeter wavelengths -- there is a gap of measurements of the solar radius as well as other parameters of the solar atmosphere. As the observational wavelength changes, the radius varies because the altitude of the dominant electromagnetic radiation is produced at different heights in the solar atmosphere. Moreover, radius variations throughout long time series are indicative of changes in the solar atmosphere that may be related to the solar cycle. Therefore, the solar radius is an important parameter for the calibration of solar atmospheric models enabling a better understanding of the atmospheric structure. In this work we use data from the Solar Submillimeter-wave Telescope (SST) and from the Atacama Large Millimeter/submillimeter Array (ALMA), at the frequencies of 100, 212, 230, and 405 GHz, to measure the equatorial and polar radii of the Sun. The radii measured with extensive data from the SST agree with the radius-vs-frequency trend present in the literature. The radii derived from ALMA maps at 230 GHz also agree with the radius-vs-frequency trend, whereas the 100-GHz radii are slightly above the values reported by other authors. In addition, we analyze the equatorial and polar radius behavior over the years, by determining the correlation coefficient between solar activity and subterahertz radii time series at 212 and 405 GHz (SST). The variation of the SST-derived radii over 13 years are correlated to the solar activity when considering equatorial regions of the solar atmosphere, and anticorrelated when considering polar regions. The ALMA derived radii time series  for 100 and 230 GHz show very similar behaviors with those of SST.
\end{abstract}


\section{INTRODUCTION}
  \label{intro}

Due to technological limitations until some decades ago, only optical observations of the Sun were available. Observations at radio wavelengths began to take place after 1950 \citep{coates58}, 
and many authors had been using measurements of the solar disk size -- \textit{i.e.} the center-to-limb distance -- as ways to determine the solar disk radius (hereafter solar radius) at different radio wavelengths \citep{coates58, wrixon70, swanson73, kisliakov75, labrum78, furst79, horne81, bachurin83, pelyushenko83, wannier83, costa86, costa99, selhorst04, alissan17, menezes17, selhorst19, selhorst19b}. There are different techniques to measure the solar radius at radio frequencies, such as the determination from total solar eclipse observations \citep{kubo93, kilcik09}, and from direct observations as the inflection point method \citep{alissan17} and the the half-power method \citep{costa99, selhorst11, menezes17}.

The study of the radio solar radius provides important information about the solar atmosphere and activity cycle \citep{swanson73, costa99, menezes17, selhorst19}. Using radio measurements of the solar radius derived from eclipse and direct observations one can probe the solar atmosphere, since these measurements show the height above the photosphere at which most of the emission at determined observation frequency is generated \citep{swanson73, menezes17, selhorst19}. However, as the observation frequency changes, the height changes as well \citep{selhorst04, selhorst19, menezes17}. To determine the height above the photosphere of the radio emission, we consider the optical solar radius. At optical wavelengths the canonical value of the mean apparent solar radius is $\Rt = $\asec{959.63} corresponding to $R_\odot=6.9599\times 10^8$ m. This value has been widely used in the literature, hence we adopt it as the reference value. In this work we focus on the solar disk radius at subterahertz radio frequencies -- \textit{i.e.}, millimeter and submillimeter wavelengths.

Therefore, with observations at several frequencies, different layers of the solar atmosphere can be observed and studied. Furthermore, these parameters can be used to improve and calibrate solar atmosphere models as an input parameter and boundary condition. In other words, solar radius at radio frequencies reflect the changes in the local distribution of temperature and density of the solar atmosphere. In Table \ref{tab:authors} and Figure \ref{fig:freq_R} we compiled data of the solar radius at several radio frequencies from different authors. We note, however, that the different works use different definitions and methods for determining the solar radius.

\begin{table}
\centering
\caption{Solar radius and altitude values at different frequencies and wavelengths.}
\begin{tabular}{lrrcc}
\hline \hline
Authors                  & Wavelength & Frequency & Radius          & Altitude \\
$\;$                     &            &           & (arcsec)        & ($10^6$ m) \\
\hline
\cite{furst79}       &     1 dm &   3 GHz &    $1070\pm17$    &    $80\pm12$ \\
\cite{furst79}       &     6 cm &   5 GHz &    $1020\pm9$     &    $44\pm7$ \\
\cite{bachurin83}    &   3.3 cm &   9 GHz &     $989\pm2$     &    $21\pm1$ \\
\cite{furst79}       &   2.7 cm &  11 GHz &     $991\pm5$     &    $23\pm4$ \\
\cite{bachurin83}    &   2.3 cm &  13 GHz &     $989\pm2$     &    $21\pm1$ \\
\cite{wrixon70}      &   1.9 cm &  16 GHz &     $990\pm4$     &    $22\pm3$ \\
\cite{selhorst04}    &   1.8 cm &  17 GHz &   $976.6\pm1.5$   &  $12.3\pm1.1$ \\
\cite{costa86}       &   1.4 cm &  22 GHz &   $981.7\pm0.8$   &  $16.0\pm0.6$ \\
\cite{furst79}       &   1.2 cm &  25 GHz &     $979\pm4$     &    $14\pm3$ \\
\cite{wrixon70}      &     1 cm &  30 GHz &     $979\pm4$     &    $14\pm3$ \\
\cite{pelyushenko83} &   8.6 mm &  35 GHz &     $979\pm3$     &    $14\pm2$ \\
\cite{selhorst19b}   &   8.1 mm &  37 GHz &     $979\pm5$     &    $14\pm4$ \\
\cite{costa86}       &   6.8 mm &  44 GHz &   $978.1\pm1.3$   &  $13.4\pm0.9$ \\
\cite{costa99}       &   6.2 mm &  48 GHz &   $983.6\pm1.9$   &  $17.4\pm1.4$ \\
\cite{pelyushenko83} &   6.2 mm &  48 GHz &   $973.1\pm2.9$   &   $9.8\pm2.1$ \\
\cite{coates58}      &   4.3 mm &  70 GHz &     $969\pm5$     &     $7\pm4$ \\
\cite{kisliakov75}   &     4 mm &  74 GHz &     $967\pm4$     &     $5\pm3$ \\
\cite{swanson73}     &   3.2 mm &  94 GHz &     $972\pm5$     &     $9\pm4$ \\
\cite{alissan17}     &     3 mm & 100 GHz &   $964.1\pm4.5$   &   $3.2\pm3.3$ \\
\cite{labrum78}      &     3 mm & 100 GHz &     $966\pm1$     &   $5.0\pm0.7$ \\
\cite{selhorst19}    &     3 mm & 100 GHz &   $965.9\pm3.2$   &   $4.5\pm2.3$ \\
\cite{wannier83}     &   2.6 mm & 115 GHz &   $969.3\pm1.6$   &   $7.0\pm1.2$ \\
\cite{menezes17}     &   1.4 mm & 212 GHz &   $966.5\pm2.8$   &   $5.0\pm2.0$ \\
\cite{alissan17}     &   1.3 mm & 230 GHz &   $961.1\pm2.5$   &   $1.1\pm1.8$ \\
\cite{selhorst19}    &   1.3 mm & 230 GHz &   $961.6\pm2.1$   &   $1.4\pm1.5$ \\
\cite{horne81}       &   1.3 mm & 231 GHz &   $968.2\pm1.0$   &   $6.2\pm7.3$ \\
\cite{menezes17}     &   0.7 mm & 405 GHz &   $966.5\pm2.7$   &   $5.0\pm2.0$ \\
\hline
\end{tabular}
\label{tab:authors} \end{table}

Another aspect to be considered is that the Sun's radius measured at the same radio frequency over time shows slight variations. Temporal series of observations obtained over many years show that the radius can be modulated with the 11-year activity cycle (mid-term variations) as well as longer periods (long term variations), as suggested by \cite{rozelot18} and references therein. \cite{costa99} measured the radius from solar maps at 48 GHz taken with a 13.7-meter dish at the Pierre Kaufmann Radio Observatory (ROPK, former Itapetinga Radio Observatory, ROI), and reported an average radius of \asec{983.6} $\pm$ \asec{1.9}. In a period of 3 years (from 1990 to 1993), temporal variations were observed following the linear relation
\begin{equation}
R_{48 \GHz} = [1.029 - 0.0015(\mathrm{year} - 1990)] \; R_{\odot} \; ,
\end{equation}
which yields a total decrease of \asec{8} if extrapolated for a period of 5.5 yr (half a cycle). Considering this short period, the data suggest that the radius decreases in phase with the monthly mean sunspot number and the soft X-ray flux from GOES (Geostationary Operational Environmental Satellite). At 37 GHz using data from Mets{\"a}hovi Radio Observatory from 1989 to 2015, \cite{selhorst19b} measured a radius of \asec{979}$\pm$\asec{5} and obtained a positive correlation coefficient of 0.44 between the monthly averages of the solar radius and the solar flux at 10.7 cm. In \cite{selhorst04}, the average solar radius from NoRH (Nobeyama Radioheliograph) daily solar maps at 17 GHz is found to be \asec{976.6} $\pm$ \asec{1.5}. Over 11 years (one solar cycle), from 1992 to 2003, the variation in the solar radius is correlated with the sunspot cycle with a coefficient $\rho=0.88$. However, the polar radius -- measurements above \dgr{60} N and below \dgr{60} S of the solar disk -- is anticorrelated with the sunspot cycle. The anticorrelation between polar radius and sunspot number yields a coefficient $\rho=-0.64$. 

At subterahertz frequencies there is a gap of measurements of the radius and other parameters of the solar atmosphere (Table \ref{tab:authors} and Figure \ref{fig:freq_R}). In this work we determine the mean equatorial and mean polar radii of the Sun at 100, 212, 230, and 405 GHz using single-dish observations from the Solar Submillimeter-wave Telescope (SST) and the Atacama Large Millimeter/submillimeter Array (ALMA).

Moreover, we present an investigation about the relation between subterahertz radii time series and solar activity cycle proxies -- solar flux at 10.7 cm, $\Fcm$, and mean magnetic field, $\Bfd$, of the Sun. We analyze the equatorial and polar radius behavior over time, with qualitative analysis for 100 and 230 GHz (ALMA) from 2015 to 2018. Also, we determine the correlation coefficient between the solar activity proxies and equatorial and polar radii time series at 212 and 405 GHz (SST), from 2007 to 2019. The methodology used for that is an improved version of the methodology presented in \cite{menezes17} with a new approach and different analysis. In this work we focused on the behavior of polar and equatorial solar radius over time.

\section{OBSERVATIONS AND DATA} \label{method}

The data used for the determination of the solar radii were provided by daily solar observations of the Solar Submillimeter-wave Telescope \citep{kaufmann08} at 212, and 405 GHz, and the Atacama Large Millimeter/submillimeter Array \citep{wootten09} at 100 and 230 GHz. The SST radio telescope was conceived to monitor continuously the submillimeter spectrum of the solar emission in quiescent and explosive conditions. In operation since 1999, the instrument is located at CASLEO observatory (lat.: \dgrms[31]{47}{54.7} S; lon.: \dgrms[69]{17}{44.1} W), at 2552 m elevation, in the Argentine Andes. From its six receivers, we used data from one radiometer at 405 GHz and three at 212 GHz, which have nominal half-power beam widths, HPBW, of \amin{2} and \amin{4} (arcminutes), respectively \citep{kaufmann08}. 

\begin{figure}
\centering
\includegraphics[width=.6\textwidth]{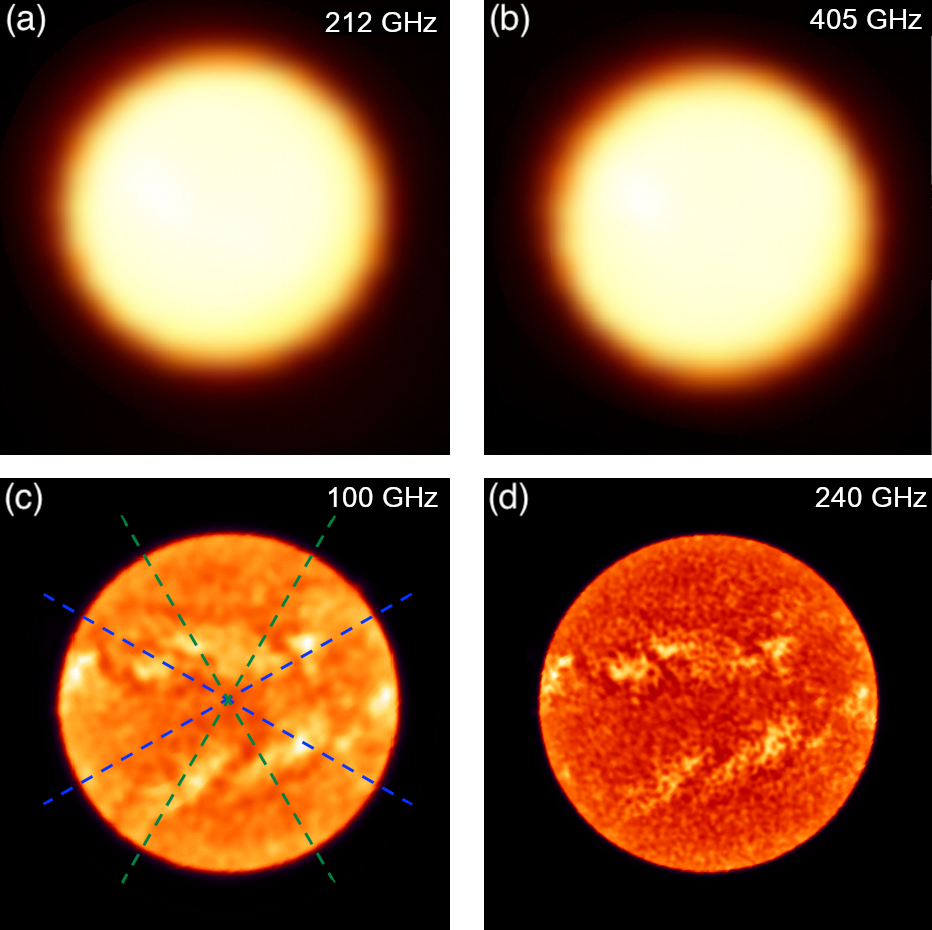}
\caption{SST maps (upper panels) obtained on 2008-02-09 and ALMA maps (lower panels) obtained on 2015-12-17. (a) 212 GHz; (b) 405 GHz; (c) 100 GHz (green lines and crosses are the constraints of the polar region and in blue, the equatorial ones); (d) 230 GHz.}
\label{fig:maps} \end{figure}

The SST radiometers were upgraded in 2006 to improve bandwidth, noise and performance, and in 2007 the SST reflector was repaired to provide better antenna efficiency, according to \cite{kaufmann08}. Therefore, we used data from 2007 onward. From 2007 to 2019 there were 3093 days of solar observation, with an average of approximately 17 solar maps per day considering all receivers, resulting in an extensive data set of 36 034 maps (27 109 at 212 GHz and 8925 at 405 GHz). Most of the maps are obtained from azimuth and elevation scans from a \amin{60}$\times$ \amin{60} area, with a \amin{2} separation between scans and tracking speed in the range of \dgr{0.1} and \dgr{0.2} per second \citep{gimenez20}. For an integration of 0.04 second, for example, it results in rectangular pixels of \amin{0.48}$\times$ \amin{2} which are then interpolated to obtain a square matrix (600$\times$ 600) as shown in the top panels of Figure \ref{fig:maps}.

ALMA is an international radio interferometre located in the Atacama Desert of northern Chile (lat.: \dgrms[23]{1}{44.40} S; lon.: \dgrms[67]{45}{18.00} W), at 5000 m elevation. From four solar single-dish observation campaigns between 2015 and 2018, we use 196 fast-scan maps (125 at 100 and 71 at 230 GHz). These maps are derived from full-disk solar observations, which consist of a circular field of view of diameter \asec{2400} using a ``double-circle'' scanning pattern \citep{white17}. The nominal spatial resolutions, HPBW, are \asec{58} and \asec{25} at 100 (Band 3) and 230 GHz (Band 6), respectively. Figure \ref{fig:maps}, bottom panels, shows examples of the ALMA maps obtained at 100 GHz (left panel) and 230 GHz (right panel), where the color represents the brightness temperature.

\subsection{Radius Determination} \label{sec:rad_determ}

Two widely used methods for measuring the radio solar radius are the inflection point method \citep{alissan17} and the the half-power method \citep{costa99, selhorst11, menezes17}. In \cite{menezes21b}, both methods are compared and it is shown how the combination of limb brightening of the solar disk with the radio-telescope beam width and shape can affect the radius determination depending on the method. \cite{menezes21b} showed that the inflection point method is less susceptible to the irregularities of the telescope beams and to the variations of the brightness temperature profiles of the Sun (\textit{e. g.} limb brightening level and active regions). Thus, the inflection point method brings a low bias to the calculation of the solar radius and, therefore, we use this method to determine the solar radius.

\begin{figure}
\centering
\includegraphics[width=.7\textwidth]{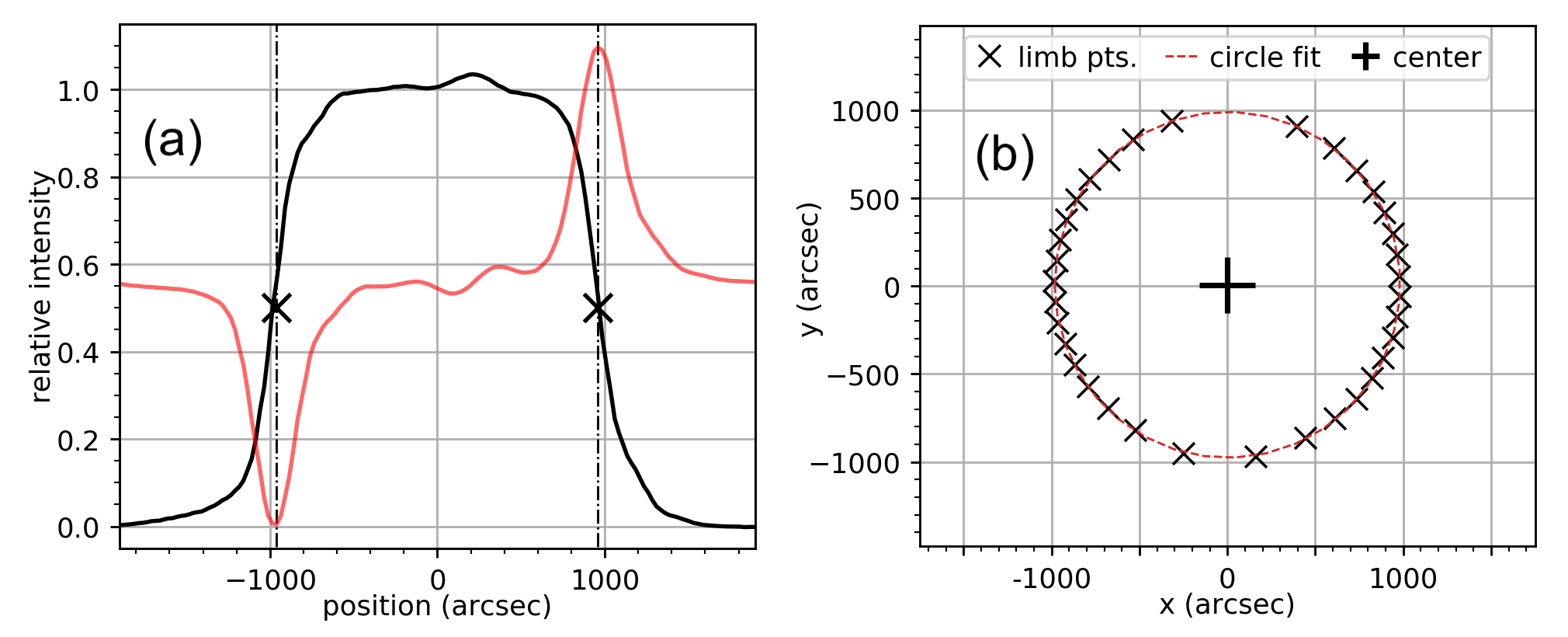}
\caption{Steps of solar radius measurement; (a) minimum and maximum points of the numerical differentiation of a scan (red curve) corresponding to the limb coordinates (black crosses); (b) limb coordinates (black crosses) extracted from a solar map with a circle fit (red dashed line).}
\label{fig:method} \end{figure}

The first step is to extract the solar limb coordinates from each map which are defined as the maximum and minimum points of the numerical differentiation red curve in Figure \ref{fig:method}-a of each scan that the telescope makes on the solar disk. All solar maps are rotated so that the position of the solar North points upwards, and the coordinates are corrected according to the eccentricity of the Earth's orbit -- the apparent radius of the Sun varies between \asec{975.3 } (perihelion) and \asec{943.2} (aphelion) during the year. During the limb points extraction, some criteria are adopted to avoid extracting limb points associated with active regions, instrumental errors or high atmospheric opacity, which may increase the calculated local radius in that region and hence the average radius. With this radius determination filter, only points with a center--to--limb distance between \asec{815} ($0.85 R_{\theta}$) and \asec{1100} ($1.15 R_{\theta}$) are considered.

The second step is to calculate the average radius of each map. The solar limb coordinates are fit by a circle and the radius is calculated as the average of the center-to-limb distances. Successive circle fits are made until certain conditions are met. For each fit, the points with center-to-limb distance (obtained from the fitting) outside the interval ($\bar{R}-$\asec{10}, $\bar{R}+$\asec{10}) are discarded, and then a new circle fit is performed with the remaining points. This process is repeated until there are at least 35\% of the  points, usually 6 out of 16 depending on the map and the latitude region -- polar or equatorial. If there are fewer points remaining, the entire map is discarded; otherwise, the radius is calculated. If the radius value is between \asec{800} and \asec{1300} and the standard deviation is below \asec{20}, then the calculated radius is stored and the next map is submitted to this process. The same method is applied to both telescopes, so that a comparison between the 212 GHz radius of SST and the 230 GHz radius of ALMA could be made.

As mentioned in Section~\ref{intro}, both the equatorial and polar radii are calculated. First, the radius of each map is determined using only the equatorial latitudes of the solar disk -- points between \dgr{30} N and \dgr{30} S. Then, the polar radius is determined considering only points above \dgr{60} N and below \dgr{60} S. These latitude boundaries are depicted in Figure \ref{fig:maps}-c, in dashed blue (equatorial) and green (polar) lines. Finally, the visible solar radius is subtracted from the the mean subterahertz radii to determine the altitudes in the atmosphere at which the 100, 212, 230, and 405 GHz emissions are predominantly produced.

Even with the strict criteria adopted in the mean radius determination, there is still a large scattering in the distribution of SST radius values. Thus, we applied a sigma clipping on the distribution, subtracting from the distribution a running mean of 300 points, and then discarding values that are outside the $\pm 2.5\sigma$ range. From the remaining values, we obtain the average radii. 

\cite{gimenez20} have shown the influence of the SST irregular beams in the study of quiescent solar structures. To assess the quality of the radius determinations, here we carry out a series of simulations convolving 2-D antenna beam matrix representations with a solar disk with uniform temperature. The results show that azimuth-elevation maps increase the uncertainty of radius determination in the direction perpendicular to scans. Since SST has an altazimuthal mount and maps are obtained at different times of the day, the uncertainty is uniformly distributed in the data set.

\subsection{Observational Time Series}

To analyze the radius variation over the solar activity cycle, we use the average taken every 6 months from the solar radii and solar proxies. We use radius daily averages, smoothed over a 100-day period to build time series at radio frequencies. As solar activity proxies, we used the 10.7-centimeter solar flux, $\Fcm$, and the intensity of the mean photospheric magnetic field, $\Bfd$, both smoothed by a 396-day running mean (13 months) to avoid the influence of annual modulations. Daily flux values of the 10.7-centimeter solar radio emission (Dominion Radio Astrophysical Observatory, DRAO), given in solar flux units ($\mathrm{1\;SFU = 10^{-22} W m^{-2} Hz^{-1}}$), is a very good proxy for solar activity cycle, as it is always measured by the same instruments, and has a smaller intrinsic scatter than the sunspot number. Another proxy used was the mean solar magnetic field, given in $\mu{T}$, provided by The Wilcox Solar Observatory \citep[WSO; ][]{scherrer77}.

\section{RESULTS} \label{result}

We used 29 088 SST maps (from a total of 36 034 maps) to calculate the radii at 212 GHz (24 186 maps) and 405 GHz (4902 maps), and 196 ALMA maps to calculate the radii at 100 GHz (125 maps) and 230 GHz (71) maps, with the method described in Section~\ref{method}.

\subsection{Subterahertz Radii}

The average equatorial and polar radii are calculated at 100, 212, 230, and 405 GHz. Moreover, the corresponding height with respect to the photosphere is deduced from the angular radii. In Table~\ref{tab:radii} the radii are listed by frequency and latitude. 

\begin{table}
\centering
\caption{Measured average radii and altitudes at subterahertz frequencies and radii derived from SSC model}
\begin{tabular}{ccccccc}
\hline \hline
Frequency & Latitude & Radius   & Radius      & Altitude      & SSC radius \\
(GHz)     &          & (arcsec) & ($10^3$ km) & ($10^{3}$ km) & (arcsec)\\
\hline \hline
\mr{2}{*}{100}
 & Equatorial &   968$\pm$3   & 702.1$\pm$2.2 & 6.1$\pm$2.2 & \mr{2}{*}{964.2} \\
 & Polar      & 968.4$\pm$2.3 & 702.4$\pm$1.7 & 6.4$\pm$1.7 & \\
\hline
\mr{2}{*}{212}
 & Equatorial &   963$\pm$4   &   699$\pm$3   &   3$\pm$3   & \mr{2}{*}{963.8} \\
 & Polar      &   963$\pm$4   &   698$\pm$3   &   2$\pm$3   & \\
\hline
\mr{2}{*}{230}
 & Equatorial & 963.7$\pm$1.8 & 698.9$\pm$1.3 & 3.0$\pm$1.3 & \mr{2}{*}{963.2} \\
 & Polar      & 963.7$\pm$1.6 & 698.9$\pm$1.2 & 3.0$\pm$1.2 & \\
\hline
\mr{2}{*}{405}
 & Equatorial &   963$\pm$5   &   699$\pm$4   &   3$\pm$4   & \mr{2}{*}{962.8} \\
 & Polar      &   963$\pm$6   &   698$\pm$4   &   2$\pm$4   & \\
\hline
\end{tabular} \label{tab:radii} \end{table}

\begin{figure}
\centering
\includegraphics[width=.7\textwidth]{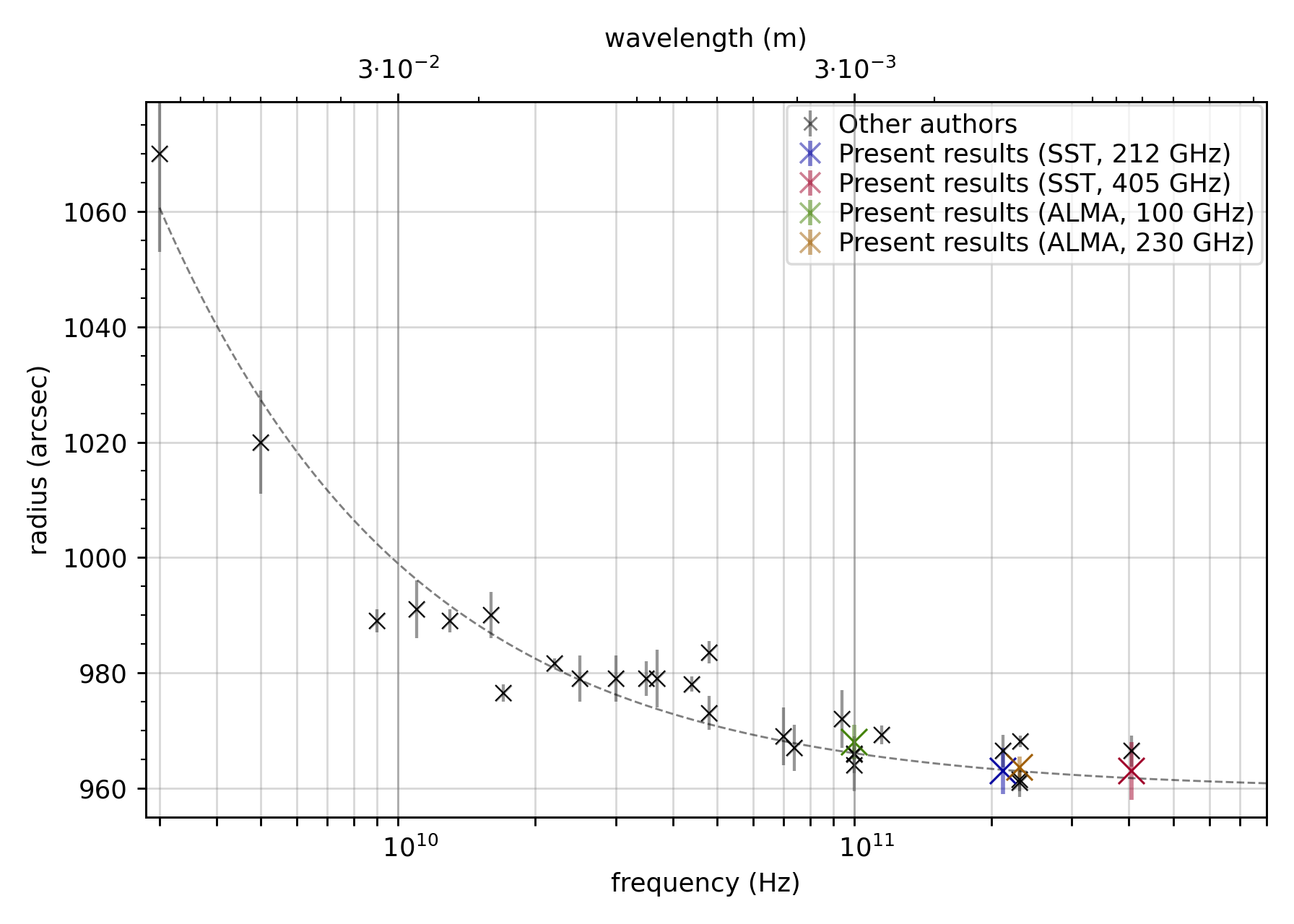}
\caption{Solar radius as a function of frequency or wavelength. The dashed line represents the exponential trend of the radius. The black crosses are previous measurements from other authors (listed in Table \ref{tab:authors}), the blue (212 GHz) and red crosses (405 GHz) are the present radius values derived from SST maps, the green (100 GHz) and yellow (230 GHz) crosses are the radii derived from ALMA.}
\label{fig:freq_R} \end{figure}

Our results are plotted with those from other authors (listed in Table \ref{tab:authors}) in Figure \ref{fig:freq_R}. To guide the eye, an exponential curve (dashed line) is over plotted to show the trend of the radius as a function of the observing frequency or wavelength, indicating that the radius decreases exponentially at radio frequencies. Note that the trend curve is just a least-square exponential fit, not a physical model. Our results are shown with green (100 GHz), blue (212 GHz), yellow (230 GHz) and red (405 GHz) crosses. SST and ALMA radii seem to agree within uncertainties with the trend of Figure~\ref{fig:freq_R} and the solar atmospheric model predictions.

Next, we used a 2-D solar atmospheric model developed by \cite{selhorst05} (hereafter referred to as the SSC model) to generate profiles of temperature brightness, $T_B$, at 100, 212, 230, and 405 GHz, which yield radii of \asec{964.2}, \asec{963.8}, \asec{963.2}, and \asec{962.8} respectively (also listed in the last column of Table \ref{tab:radii}). Our results at 212, 230 and 405 GHz are very close to the radii derived from the model, whereas the 100-GHz radii are about \asec{4} bigger, not agreeing with the model. 

\subsection{Correlation with Solar Activity}

\begin{figure}
\centering
\includegraphics[trim=0pt 22pt 0pt 22pt, clip=true, width=.82\textwidth ]{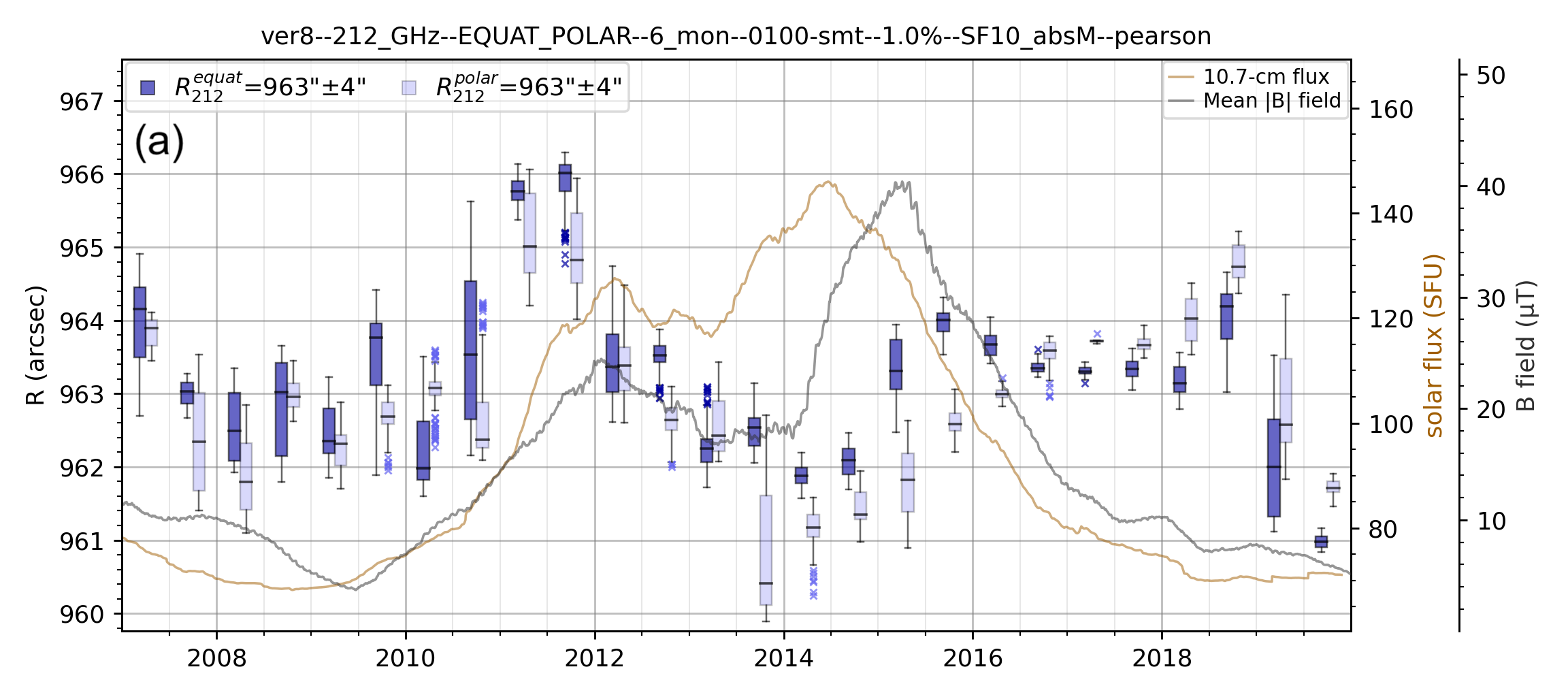}
\includegraphics[trim=0pt 6pt 0pt 22pt, clip=true, width=.82\textwidth ]{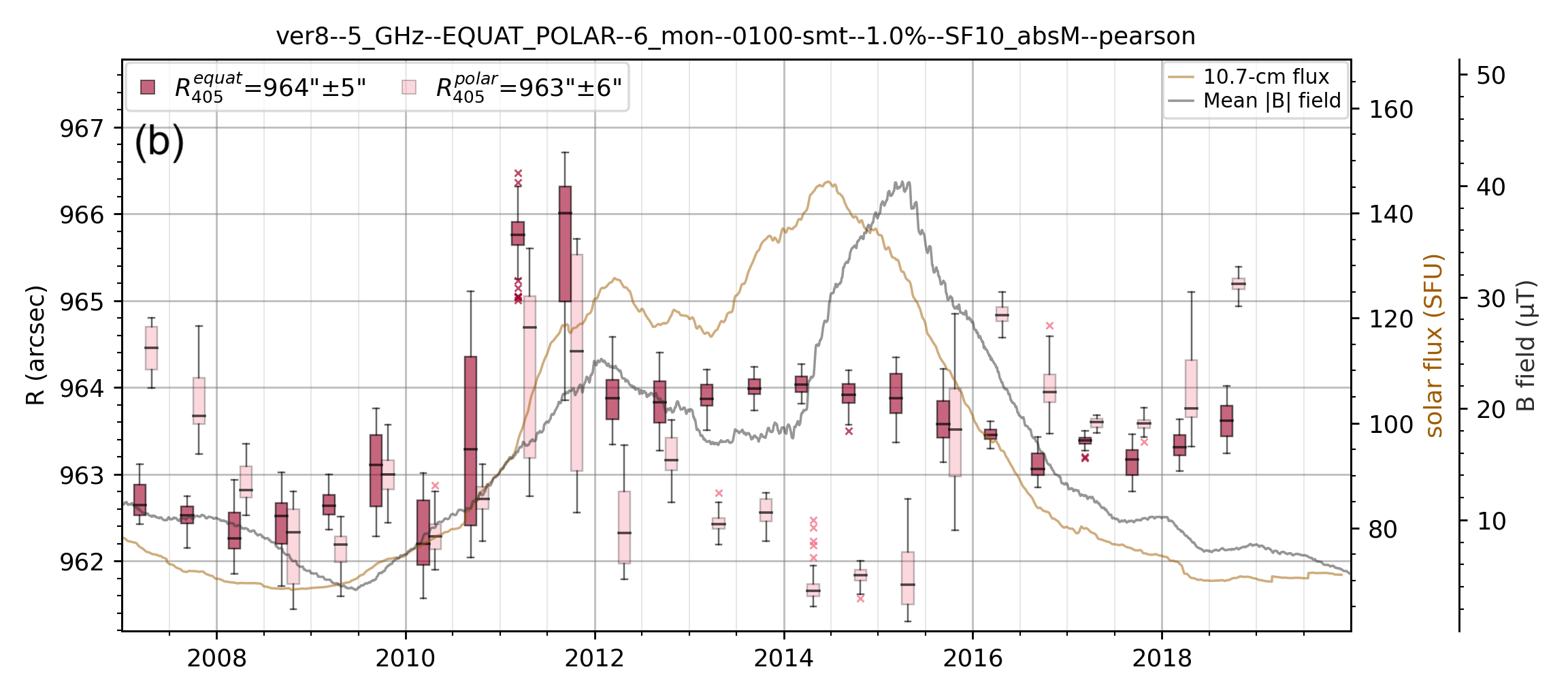}
\caption{
Solar radius time series, $\Rt$, at 212 and 405 GHz from January 2007 to December 2019. The top panel (a) show time series for equatorial (dark blue) and polar (light blue) radii at 212 GHz. The bottom panel (b) show time series for equatorial (red) and polar (pink) radii at 405 GHz. The orange and gray lines represent $\Fcm$ and $\Bfd$, respectively. Every box represents data for a 6-month period: the horizontal line inside the box is the period median and the crosses represent the outliers of the distribution.}
\label{fig:timeseries} \end{figure}

We investigated the temporal variation of the solar radius and its relationship with the 11-year solar cycle. The subterahertz radius time series was analyzed from 2007 to 2016 using radii derived from the whole solar disk only \citep{menezes17}. Here, we analyze a 13-year period -- from January 2007 to December 2019 -- using radii derived from equatorial and polar latitude regions. As solar activity proxies, we used the 10.7-cm solar flux, $\Fcm$, and the mean solar magnetic field intensity, $\Bfd$. The results are plotted in Figure \ref{fig:timeseries}.

\begin{figure}
\centering
\includegraphics[trim=0pt 8pt 0pt 8pt, clip=true, width=.78\textwidth ]{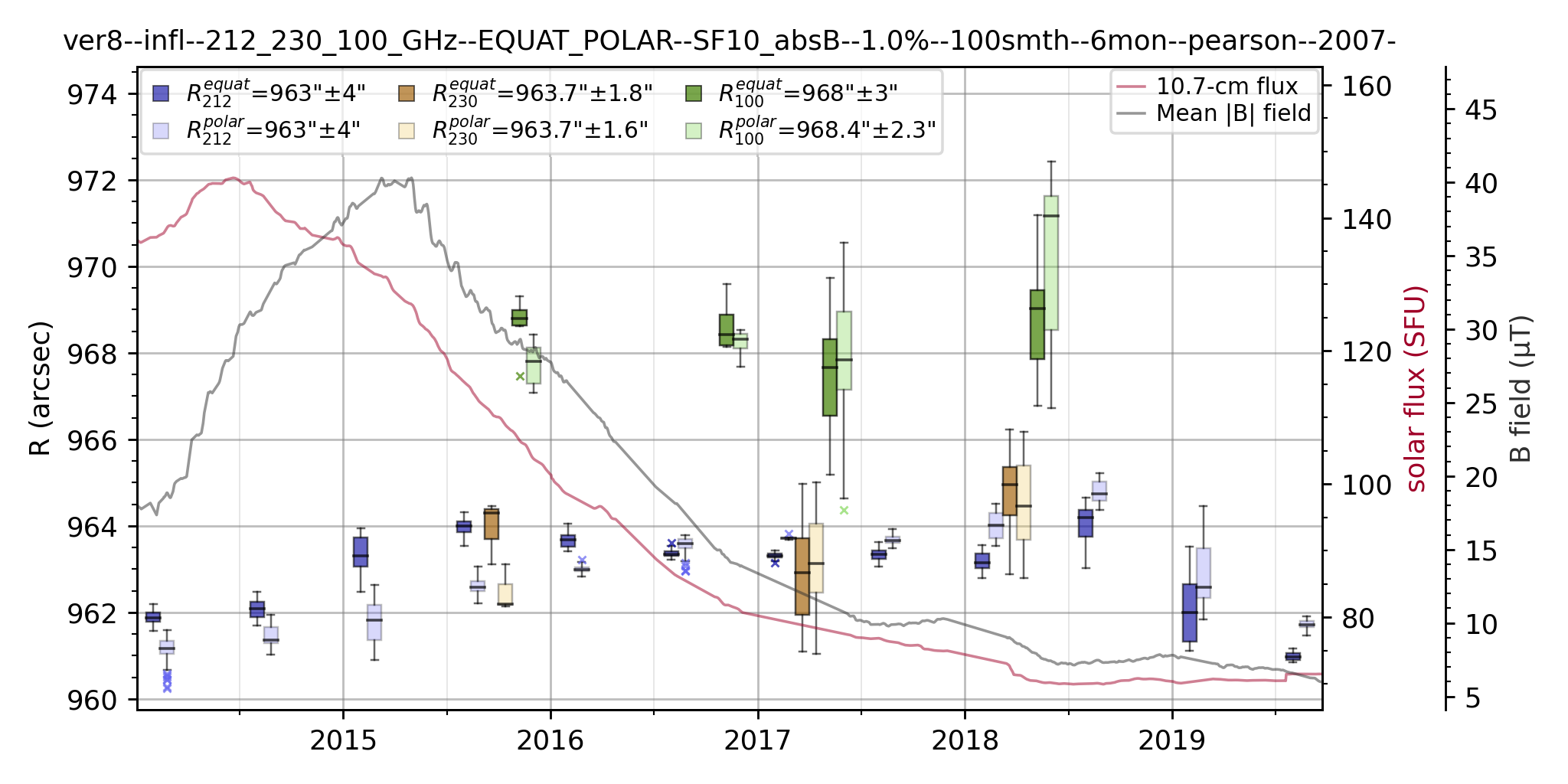}
\caption{
Boxplot time series for equatorial $\Rt$ at 100 (green), 212 (dark blue) and 230 GHz (orange), and for polar $\Rt$ at 100 (light green), 212 (light blue) and 230 GHz (yellow), from January 2014 to December 2019. Every box represents data for a 6-month period: the horizontal line inside the box is the period median. The dots represent the outliers of the distribution. The red line represents $\Fcm$ and the gray line represents $\Bfd$.}
\label{fig:timeser_S_A} \end{figure}

We compared the equatorial and polar radius time series at 212 GHz (SST) with the time series of the equatorial and polar radii at 230 GHz (ALMA), with 6-month averages. The radii derived from SST's maps are found to have both average values and behavior over time very close to ALMA's, which can be seen in Figure \ref{fig:timeser_S_A}. Moreover, besides the lower frequency and higher radius values, equatorial and polar radii at 100 GHz seem to have similar behavior with those of 212 and 230 GHz time series.

The comparison of the solar radii in time with the solar proxies is summarized in Table~\ref{tab:correl}, where the calculated correlation coefficients, $\rho$, are organized by frequency, solar latitude regions and solar proxies. 
The correlation coefficients between equatorial radius, $\Req$, and solar proxies are very low (0.05 and 0.16, respectively). However, for $\Rpo$ the coefficients indicate weak anticorrelation (-0.36 for $\Fcm$ and -0.23 for $\Bfd$). $\Reqq$ is moderately correlated with $\Fcm$ (0.64) and $\Bfd$ (0.50), while $\Rpoo$ is weakly anticorrelated with $\Fcm$ (-0.39) and $\Bfd$ (-0.23).
In summary, the radii -- $\Req$, $\Rpo$, $\Reqq$, and $\Rpoo$ -- have stronger correlation with $\Bfd$ than with $\Fcm$.

\begin{table}
\centering
\caption{Linear correlation coefficients, $\rho$, between solar radii and solar proxies.}
\begin{tabular}{lccc}
\hline \hline
Frequency & Latitude & $\rho_{\Fcm}$ & $\rho_{\Bfd}$ \\
\hline
212 GHz & Equatorial &        0.05   &         0.16 \\ 
212 GHz & Polar      &       -0.36   &        -0.23 \\ 
405 GHz & Equatorial &        0.64   &         0.50 \\ 
405 GHz & Polar      &       -0.39   &        -0.23 \\ 
\hline
\end{tabular} \label{tab:correl} \end{table}

\section{DISCUSSION AND CONCLUSIONS} \label{conclusion}

From the extensive SST and the ALMA data set we determined the polar and equatorial radii of the Sun at 100, 212, 230, and 405 GHz. The average radii are in agreement with the radius-vs-frequency trend (Fig. \ref{fig:freq_R}), however the values obtained for the ALMA maps are higher than those obtained by \cite{selhorst19} and \cite{alissan17}. The 100-GHz average radius is about \asec{4} larger then that measured by \cite{alissan17}, and about \asec{2} larger then those measured by \cite{labrum78} and \cite{selhorst19}. Nevertheless, the 100-GHz radius was measured by \cite{alissan17} and \cite{selhorst19} using maps observed only in December 2015, and by \cite{labrum78} using observations of the total eclipse of 1976 October 23. Also there is a difference of roughly \asec{4} between our 100-GHz measurements and the radius derived from the SSC model. \cite{menezes21b} showed that the radius increases as a function of the limb brightening intensity. By convolving the ALMA beam with the SSC model (limb brightening level 33.6\% above the quiet Sun level), they estimated an increase of \asec{2.3} on the radius at 100 GHz. The limb brightening levels could be increasing over time as the solar activity decreases, and therefore affecting the solar radius.



Moreover, we have analyzed the subterahertz solar radius time series for more than a solar cycle, over 13 years (2007 -- 2019), at 212 and 405 GHz. The radii time series are not strongly correlated with the solar proxies, however one can observe particular behaviors of these time series, with the polar radius being anticorrelated and the equatorial radius being correlated with $\Bfd$. This is a similar behavior of the radio radius presented in the literature for lower frequencies \cite{costa99, selhorst04, selhorst19b}, which are expected to be correlated to the solar cycle -- positively for the average radius and negatively for the polar radius. The equatorial radii time series are expected to be positively correlated to the solar cycle, since the equatorial regions are more affected by the increase of active regions during solar maxima, making the solar atmosphere warmer in these regions. On the other hand, the anticorrelation between polar radius time series and the solar activity proxies could be explained by a possible increase of polar limb brightening during solar minima. In a previous study of polar limb brightening seen in 17 GHz solar maps, \cite{selhorst03} concluded that the intensity of this brightening was anticorrelated with the solar cycle.

The 212 and 405-GHz radius time series do not seem to be as well defined as at lower frequencies such as 17, 37, and 48 GHz \citep{costa99, selhorst04, selhorst19b}. The subterahertz radii measured in this work correspond to emission altitudes much closer to the photosphere. Therefore, our results  correspond to a frequency range (212 to 405 GHz) that probably reflects the behavior from both the photosphere and the lower chromosphere. In part, this would explain the weak (212 GHz) and the moderate (405 GHz) correlation  of the radii with the solar proxies.

Moreover, we could also observe analogous behavior between SST and ALMA time series, even with the shorter period of ALMA solar observations. Besides lower frequency and higher radius values, equatorial and polar radii at 100 GHz seem to have similar variations with those of 212 and 230 GHz time series. At both ALMA frequencies, the equatorial radii decrease from 2015 to 2017 and increase from 2017 to 2018 -- roughly close to $\Bfd$. The polar radii time series just increases from 2015 to 2018, which is the opposite of $\Bfd$. As the subterahertz radiation is influenced by Bremsstrahlung emission, the observed variations are expected since the solar atmosphere's density, temperature, and magnetic field change with time. Longer periods of future observations at these frequencies will reveal the polar and equatorial trends of the solar atmosphere.

Measuring the solar radius at subterahertz frequencies is not an easy task as well as to compare the values from different studies, since different instruments observe at different band widths. Our results are important to test atmospheric models, and better understand the solar cycle, since they probe directly different layers of the solar atmosphere over time. More studies of such kind at other frequencies and for longer periods of time are needed to achieve this goal.

\begin{acknowledgements}

The authors thank J. Valle, P. J. A. Sim\~oes, and D. Cornejo for fruitful discussions. F. M. thanks MackPesquisa and CAPES for the scholarship.

We acknowledge the financial support for operation of the Solar Submillimeter Telescope (SST) from S{\~a}o Paulo Research Foundation (FAPESP) Proc. \#2013/24155-3 and AFOSR Grant \#FA9550-16-1-0072. A. V. acknowledges partial financial support from the FAPESP, grant number 2013/10559-5.
C.L.S. acknowledges financial support from the FAPESP, grant number 2019/ 03301-8. C.G.G.C. is grateful to CNPq for providing support to this research (grant 307722/2019-8).

This work is based on data acquired at \textit{Complejo Astronómico El Leoncito}, operated under agreement between the \textit{Consejo Nacional de Investigaciones Cient\'ificas y T\'ecnicas de la Rep\'ublica Argentina} and the \textit{National Universities} of \textit{La Plata}, \textit{C\'ordoba} and \textit{San Ju\'an}.

This paper makes use of the following ALMA data: ADS/JAO.ALMA \#2011.0.00020.SV \#2016.1.00050.S, \#2016.1.00070.S, \#2016.1.00156.S, \#2016.1.00166.S, \#2016.1.00182.S, \#2016.1.00201.S, \#2016.1.00202.S, \#2016.1.00423.S, \#2016.1.00572.S, \#2016.1.00788.S, \#2016.1.01129.S, \#2016.1.01408.S, \#2016.1.01532.S, \#2017.1.00009.S, \#2017.1.00870.S, \#2017.1.01138.S. ALMA is a partnership of ESO (representing its member states), NSF (USA) and NINS (Japan), together with NRC (Canada), MOST and ASIAA (Taiwan), and KASI (Republic of Korea), in cooperation with the Republic of Chile. The Joint ALMA Observatory is operated by ESO, AUI/NRAO and NAOJ. 

This work makes use of data provided by The Wilcox Solar Observatory, which was obtained via the web site \url{http://wso.stanford.edu}.

Also, we acknowledge Dr. Ken Tapping, Research Council Officer, DRAO, and the Solar Radio Monitoring Program, operated jointly by the National Research Council Canada and Natural Resources Canada, for the radio flux data.
\end{acknowledgements}

\bibliography{Main.bib}

\begin{thebibliography}{}
\expandafter\ifx\csname natexlab\endcsname\relax\def\natexlab#1{#1}\fi
\providecommand{\url}[1]{\href{#1}{#1}}
\providecommand{\dodoi}[1]{doi:~\href{http://doi.org/#1}{\nolinkurl{#1}}}
\providecommand{\doeprint}[1]{\href{http://ascl.net/#1}{\nolinkurl{http://ascl.net/#1}}}
\providecommand{\doarXiv}[1]{\href{https://arxiv.org/abs/#1}{\nolinkurl{https://arxiv.org/abs/#1}}}

\bibitem[{{Alissandrakis} {et~al.}(2017){Alissandrakis}, {Patsourakos},
  {Nindos}, \& {Bastian}}]{alissan17}
{Alissandrakis}, C.~E., {Patsourakos}, S., {Nindos}, A., \& {Bastian}, T.~S.
  2017, \aap, 605, A78, \dodoi{10.1051/0004-6361/201730953}

\bibitem[{{Bachurin}(1983)}]{bachurin83}
{Bachurin}, A.~F. 1983, Izvestiya Ordena Trudovogo Krasnogo Znameni Krymskoj
  Astrofizicheskoj Observatorii, 68, 68

\bibitem[{{Coates}(1958)}]{coates58}
{Coates}, R.~J. 1958, \apj, 128, 83, \dodoi{10.1086/146518}

\bibitem[{{Costa} {et~al.}(1986){Costa}, {Homor}, \& {Kaufmann}}]{costa86}
{Costa}, J.~E.~R., {Homor}, J.~L., \& {Kaufmann}, P. 1986, in Solar Flares and
  Coronal Physics Using P/OF as a Research Tool, ed. E.~{Tandberg}, R.~M.
  {Wilson}, \& R.~M. {Hudson}

\bibitem[{{Costa} {et~al.}(1999){Costa}, {Silva}, {Makhmutov}, {Rolli},
  {Kaufmann}, \& {Magun}}]{costa99}
{Costa}, J.~E.~R., {Silva}, A.~V.~R., {Makhmutov}, V.~S., {et~al.} 1999, \apjl,
  520, L63, \dodoi{10.1086/312132}

\bibitem[{{F{\"u}rst} {et~al.}(1979){F{\"u}rst}, {Hirth}, \&
  {Lantos}}]{furst79}
{F{\"u}rst}, E., {Hirth}, W., \& {Lantos}, P. 1979, \solphys, 63, 257,
  \dodoi{10.1007/BF00174532}

\bibitem[{{Gim{\'e}nez de Castro} {et~al.}(2020){Gim{\'e}nez de Castro},
  {Pereira}, {Valle Silva}, {Selhorst}, {Mand rini}, {Cristiani}, {Raulin}, \&
  {Valio}}]{gimenez20}
{Gim{\'e}nez de Castro}, C.~G., {Pereira}, A. L.~G., {Valle Silva}, J.~F.,
  {et~al.} 2020, arXiv e-prints, arXiv:2009.03445.
\newblock \doarXiv{2009.03445}

\bibitem[{{Horne} {et~al.}(1981){Horne}, {Hurford}, {Zirin}, \& {de
  Graauw}}]{horne81}
{Horne}, K., {Hurford}, G.~J., {Zirin}, H., \& {de Graauw}, T. 1981, \apj, 244,
  340, \dodoi{10.1086/158711}

\bibitem[{{Kaufmann} {et~al.}(2008){Kaufmann}, {Levato}, {Cassiano}, {Correia},
  {Costa}, {Gim{\'e}nez de Castro}, {Godoy}, {Kingsley}, {Kingsley}, {Kudaka},
  {Marcon}, {Martin}, {Marun}, {Melo}, {Pereyra}, {Raulin}, {Rose}, {Silva
  Valio}, {Walber}, {Wallace}, {Yakubovich}, \& {Zakia}}]{kaufmann08}
{Kaufmann}, P., {Levato}, H., {Cassiano}, M.~M., {et~al.} 2008, in
  \textit{Proc.} SPIE, Vol. 7012, Ground-based and Airborne Telescopes II,
  70120L, \dodoi{10.1117/12.788889}

\bibitem[{Kilcik {et~al.}(2009)Kilcik, Sigismondi, Rozelot, \& Guhl}]{kilcik09}
Kilcik, A., Sigismondi, C., Rozelot, J., \& Guhl, K. 2009, Solar Physics, 257,
  237

\bibitem[{{Kislyakov} {et~al.}(1975){Kislyakov}, {Kulikov}, {Fedoseev}, \&
  {Chernyshev}}]{kisliakov75}
{Kislyakov}, A.~G., {Kulikov}, I.~I., {Fedoseev}, L.~I., \& {Chernyshev}, V.~I.
  1975, Soviet Astronomy Letters, 1, 79

\bibitem[{Kubo(1993)}]{kubo93}
Kubo, Y. 1993, Publications of the Astronomical Society of Japan, 45, 819

\bibitem[{{Labrum} {et~al.}(1978){Labrum}, {Archer}, \& {Smith}}]{labrum78}
{Labrum}, N.~R., {Archer}, J.~W., \& {Smith}, C.~J. 1978, \solphys, 59, 331,
  \dodoi{10.1007/BF00951837}

\bibitem[{{Menezes} {et~al.}(2021){Menezes}, {Selhorst}, \& {Gim{\'e}nez de
  Castro}}]{menezes21b}
{Menezes}, F., {Selhorst}, C.~L., \& {Gim{\'e}nez de Castro}, C.~G.~{Valio}, A.
  2021, [Unpublished manuscript]

\bibitem[{{Menezes} \& {Valio}(2017)}]{menezes17}
{Menezes}, F., \& {Valio}, A. 2017, \solphys, 292, 195,
  \dodoi{10.1007/s11207-017-1216-y}

\bibitem[{{Pelyushenko} \& {Chernyshev}(1983)}]{pelyushenko83}
{Pelyushenko}, S.~A., \& {Chernyshev}, V.~I. 1983, \sovast, 27, 340

\bibitem[{{Rozelot} {et~al.}(2018){Rozelot}, {Kosovichev}, \&
  {Kilcik}}]{rozelot18}
{Rozelot}, J.~P., {Kosovichev}, A.~G., \& {Kilcik}, A. 2018, Sun and Geosphere,
  13, 63.
\newblock \doarXiv{1804.06930}

\bibitem[{{Scherrer} {et~al.}(1977){Scherrer}, {Wilcox}, {Svalgaard}, {Duvall},
  {Dittmer}, \& {Gustafson}}]{scherrer77}
{Scherrer}, P.~H., {Wilcox}, J.~M., {Svalgaard}, L., {et~al.} 1977, \solphys,
  54, 353, \dodoi{10.1007/BF00159925}

\bibitem[{{Selhorst} {et~al.}(2011){Selhorst}, {Gim{\'e}nez de Castro},
  {V{\'a}lio}, {Costa}, \& {Shibasaki}}]{selhorst11}
{Selhorst}, C.~L., {Gim{\'e}nez de Castro}, C.~G., {V{\'a}lio}, A., {Costa},
  J.~E.~R., \& {Shibasaki}, K. 2011, \apj, 734, 64,
  \dodoi{10.1088/0004-637X/734/1/64}

\bibitem[{{Selhorst} {et~al.}(2019{\natexlab{a}}){Selhorst}, {Kallunki},
  {Gim{\'e}nez de Castro}, {Valio}, \& {Costa}}]{selhorst19b}
{Selhorst}, C.~L., {Kallunki}, J., {Gim{\'e}nez de Castro}, C.~G., {Valio}, A.,
  \& {Costa}, J. E.~R. 2019{\natexlab{a}}, \solphys, 294, 175,
  \dodoi{10.1007/s11207-019-1568-6}

\bibitem[{{Selhorst} {et~al.}(2004){Selhorst}, {Silva}, \&
  {Costa}}]{selhorst04}
{Selhorst}, C.~L., {Silva}, A.~V.~R., \& {Costa}, J.~E.~R. 2004, \aap, 420,
  1117, \dodoi{10.1051/0004-6361:20034382}

\bibitem[{{Selhorst} {et~al.}(2005){Selhorst}, {Silva}, \&
  {Costa}}]{selhorst05}
---. 2005, \aap, 433, 365, \dodoi{10.1051/0004-6361:20042043}

\bibitem[{{Selhorst} {et~al.}(2003){Selhorst}, {Silva}, {Costa}, \&
  {Shibasaki}}]{selhorst03}
{Selhorst}, C.~L., {Silva}, A.~V.~R., {Costa}, J.~E.~R., \& {Shibasaki}, K.
  2003, \aap, 401, 1143, \dodoi{10.1051/0004-6361:20030071}

\bibitem[{{Selhorst} {et~al.}(2019{\natexlab{b}}){Selhorst}, {Sim{\~o}es},
  {Braj{\v{s}}a}, {Valio}, {Gim{\'e}nez de Castro}, {Costa}, {Menezes},
  {Rozelot}, {Hales}, {Iwai}, \& {White}}]{selhorst19}
{Selhorst}, C.~L., {Sim{\~o}es}, P. J.~A., {Braj{\v{s}}a}, R., {et~al.}
  2019{\natexlab{b}}, \apj, 871, 45, \dodoi{10.3847/1538-4357/aaf4f2}

\bibitem[{{Swanson}(1973)}]{swanson73}
{Swanson}, P.~N. 1973, \solphys, 32, 77, \dodoi{10.1007/BF00152730}

\bibitem[{{Wannier} {et~al.}(1983){Wannier}, {Hurford}, \&
  {Seielstad}}]{wannier83}
{Wannier}, P.~G., {Hurford}, G.~J., \& {Seielstad}, G.~A. 1983, \apj, 264, 660,
  \dodoi{10.1086/160639}

\bibitem[{{White} {et~al.}(2017){White}, {Iwai}, {Phillips}, {Hills}, {Hirota},
  {Yagoubov}, {Siringo}, {Shimojo}, {Bastian}, {Hales}, {Sawada}, {Asayama},
  {Sugimoto}, {Marson}, {Kawasaki}, {Muller}, {Nakazato}, {Sugimoto}, {Braj{\v
  s}a}, {Skoki{\'c}}, {B{\'a}rta}, {Kim}, {Remijan}, {de Gregorio}, {Corder},
  {Hudson}, {Loukitcheva}, {Chen}, {De Pontieu}, {Fleishmann}, {Gary},
  {Kobelski}, {Wedemeyer}, \& {Yan}}]{white17}
{White}, S.~M., {Iwai}, K., {Phillips}, N.~M., {et~al.} 2017, \solphys, 292,
  88, \dodoi{10.1007/s11207-017-1123-2}

\bibitem[{{Wootten} \& {Thompson}(2009)}]{wootten09}
{Wootten}, A., \& {Thompson}, A.~R. 2009, IEEE Proceedings, 97, 1463,
  \dodoi{10.1109/JPROC.2009.2020572}

\bibitem[{{Wrixon}(1970)}]{wrixon70}
{Wrixon}, G.~T. 1970, \nat, 227, 1231, \dodoi{10.1038/2271231a0}

\end{thebibliography}
\bibliographystyle{aasjournal}

\end{document}